\documentclass[preprint,lettersize,journal,twoside,web]{IEEEtran}
\usepackage{booktabs}
\usepackage{cite}
\usepackage{amsmath,amssymb,amsfonts}
\usepackage{graphicx}
\usepackage{multirow}
\usepackage{color}
\usepackage{soul}

\usepackage[switch]{lineno}

\usepackage{algorithmic}
\usepackage{algorithm}
\usepackage{array}
\usepackage[caption=false,font=normalsize,labelfont=sf,textfont=sf]{subfig}
\usepackage{placeins}
\usepackage{stfloats}
\usepackage{url}
\usepackage{textcomp}
\usepackage{verbatim}
\usepackage{graphicx}

\usepackage{bm}
\begin{document}
	\title{\Huge SWDL: \underline{S}tratum-\underline{W}ise \underline{D}ifference \underline{L}earning with Deep Laplacian Pyramid for Semi-Supervised 3D Intracranial Hemorrhage Segmentation}
	\author{Cheng Wang, Siqi Chen, Donghua Mi, Yang Chen, Yudong Zhang, Yinsheng Li
	\thanks{
	Scientific correspondence should be addressed to Y.~Zhang: yudongzhang@ieee.org, Y.~chen: chenyang.list@seu.edu.cn, Y.~Li: ys.li2@siat.ac.cn and D.~Mi: midonghua131@163.com.} 
    \thanks{C.~Wang (email: chengwang0711@163.com) is with the School of Computer Science and Engineering, Southeast University, Nanjing 210009, China, and also with Research Center for Medical Artificial Intelligence, Shenzhen Institutes of Advanced Technology, Chinese Academy of Sciences, Shenzhen 518055, China.}
    \thanks{S.~Chen (e-mail: chen4731415926@163.com), and D.~Mi (e-mail: midonghua131@163.com) are with Beijing Tiantan Hospital, Capital Medical University, Beijing 100070, China.}
    \thanks{Y.~chen (e-mail: chenyang.list@seu.edu.cn), and Y.~Zhang (e-mail: yudongzhang@ieee.org) are with the School of Computer Science and Engineering, Southeast University, Nanjing 210009, China.}
    \thanks{Y.~Li (email: ys.li2@siat.ac.cn) is with Research Center for Medical Artificial Intelligence, Shenzhen Institutes of Advanced Technology, Chinese Academy of Sciences, Shenzhen 518055, China.}
    \thanks{This work was supported in part by the Civil Space Technology Pre-research Program (Grant No. D010101), in part by the National Natural Science Foundation of China (Grant No. 62301545), and in part by the Shenzhen Science and Technology Program (Grant No. JCYJ20220818101803008).}
	\vspace{-3em}
	}
\maketitle
\begin{abstract}
Recent advances in medical imaging have established deep learning-based segmentation as the predominant approach, though it typically requires large amounts of manually annotated data. However, obtaining annotations for intracranial hemorrhage (ICH) remains particularly challenging due to the tedious and costly labeling process. Semi-supervised learning (SSL) has emerged as a promising solution to address the scarcity of labeled data, especially in volumetric medical image segmentation. Unlike conventional SSL methods that primarily focus on high-confidence pseudo-labels or consistency regularization, we propose SWDL-Net, a novel SSL framework that exploits the complementary advantages of Laplacian pyramid and deep convolutional upsampling. The Laplacian pyramid excels at edge sharpening, while deep convolutions enhance detail precision through flexible feature mapping. Our framework achieves superior segmentation of lesion details and boundaries through a difference learning mechanism that effectively integrates these complementary approaches. Extensive experiments on a 271-case ICH dataset and public benchmarks demonstrate that SWDL-Net outperforms current state-of-the-art methods in scenarios with only 2\% labeled data. Additional evaluations on the publicly available Brain Hemorrhage Segmentation Dataset (BHSD) with 5\% labeled data further confirm the superiority of our approach. Code and data have been released at https://github.com/SIAT-CT-LAB/SWDL.
\end{abstract}


\section{Introduction}
Intracranial hemorrhage (ICH) segmentation in non-contrast CT (NCCT) scans presents a critical challenge in emergency neuroimaging, with immediate implications for stroke diagnosis and treatment planning \cite{monteiro2020multiclass}. While deep learning has revolutionized medical image analysis \cite{ronneberger2015u}, its application to ICH segmentation faces unique obstacles due to the complex appearance variations of hemorrhages and the scarcity of expert annotations \cite{arabo2022optimized}. Recent epidemiological studies underscore the urgency of addressing intracerebral hemorrhage (ICH), as it demonstrates mortality rates exceeding 50\% within one year of onset and often leaves survivors with significant functional and cognitive impairments \cite{puy2023intracerebral}.

Despite these clinical imperatives, the development of robust segmentation algorithms must contend with both technical complexities and practical data constraints. Recent years have witnessed significant advancements in medical imaging segmentation methodologies. Fully-supervised approaches have evolved from traditional U-Net architectures \cite{ronneberger2015u} to more sophisticated designs. Transformer-based models like Swin-UNet \cite{cao2023swin} has demonstrated superior capability in capturing long-range dependencies, while hybrid architectures such as nnFormer \cite{zhou2022nnformer} and MedFormer \cite{chowdary2024med} effectively combine convolutional and self-attention mechanisms. Particularly noteworthy are the 3D contextual approaches like HDC-Net \cite{wang2021hdc} and Dense-UNet \cite{cai2020dense}, which could address the volumetric nature of ICH lesions. Self-supervised pretraining strategies \cite{chen2022self} have shown promise in mitigating data scarcity, with contrastive learning frameworks like ConResNet \cite{zhang2020inter} achieving significant performance gains. However, these methods remain constrained by their dependence on large annotated datasets - a significant limitation given the expertise required for accurate ICH labeling.

Semi-supervised learning (SSL) offers a promising solution to the annotation bottleneck \cite{jiao2024learning}, achieving impressive results with limited labeled data through two primary strategies: pseudo-labeling and consistency regularization. Consistency-based methods enforce stable predictions under perturbations and include techniques such as I2CS \cite{xie2021intra} (aligning attention maps across labeled and unlabeled data), URPC \cite{luo2022semi} (uncertainty-rectified pyramid consistency), MC-Net+ \cite{wu2022mutual} (agreement among structurally diverse decoders), Diverse Co-training \cite{li2023diverse} (cross-pseudo supervision), SemiGroup \cite{li2023semi} (collaborative teacher-student groups), and UniMatch \cite{yang2023revisiting} (weak-to-strong regularization). Pseudo-labeling methods focus on generating high-quality pseudo-labels, with techniques like BoostMIS \cite{zhang2022boostmis} (adaptive thresholds), UA-MT \cite{yu2019uncertainty} (teacher-student supervision), PEFAT \cite{zeng2023pefat} (loss distribution analysis), and ComWin \cite{wu2023compete} (boundary-aware attention).

Recent semi-supervised ICH segmentation methods exhibit notable limitations. Uncertainty-guided approaches \cite{emon2025uncertainty} suffer from threshold sensitivity, while weakly supervised CAM-based methods \cite{ramananda2025label} yield incomplete boundaries. Cut-paste augmentation \cite{yap2023cut} may generate anatomically unrealistic samples, and YOLO-SAM hybrids \cite{spiegler2024weakly} inherit resolution constraints from detection models. 

We present SWDL-Net, a novel SSL framework that transforms the anisotropy challenge into a learning opportunity through differential feature analysis between Laplacian pyramid and deep convolutional upsampling paths. Our key contributions are:

\begin{itemize}
    \item A \textbf{hybrid encoder-decoder architecture} that synergizes a CNN-based encoder with dual decoders (Laplacian pyramid and deep convolutional), building upon multi-path design principles while introducing novel difference learning mechanisms between pathways.
    
    \item An innovative \textbf{skip-connected difference feature computation} that preserves critical information through a novel pathway interaction mechanism. Unlike conventional concatenation-based fusion, our approach first computes inter-pathway differences, applies nonlinear activation, and then performs weighted summation - effectively capturing complementary information between the deep laplacian pyramid and deep convolutional pathways. 
    
    \item A \textbf{stratum-adaptive pyramid optimization} mechanism that dynamically adjusts the Laplacian pyramid's depth according to input dimensions, achieving an optimal trade-off between reconstruction fidelity and computational efficiency. This approach effectively prevents information redundancy while fully leveraging the deep Laplacian pyramid's dual advantages:  superior edge information extraction and precise preservation of high-frequency details critical for medical image segmentation.
\end{itemize}

Our experimental results demonstrate consistent improvements over existing methods in both segmentation accuracy and boundary precision, while maintaining computational efficiency suitable for clinical deployment. Through extensive ablation studies and rigorous statistical analysis, we validate the framework's robustness and reliability. Comprehensive evaluations across two benchmarks - a 271-case clinical dataset and the public Brain Hemorrhage Segmentation Dataset (BHSD) \cite{wu2023bhsd} - establish \textbf{state-of-the-art (SOTA)} performance. Notably, our method achieves superior results in challenging low-label regimes (2-5\% labeled data), setting a new benchmark for semi-supervised medical image segmentation.

\section{SWDL-Net}
\begin{figure*}[htbp] 
    \centering 
    \includegraphics[width=\textwidth]{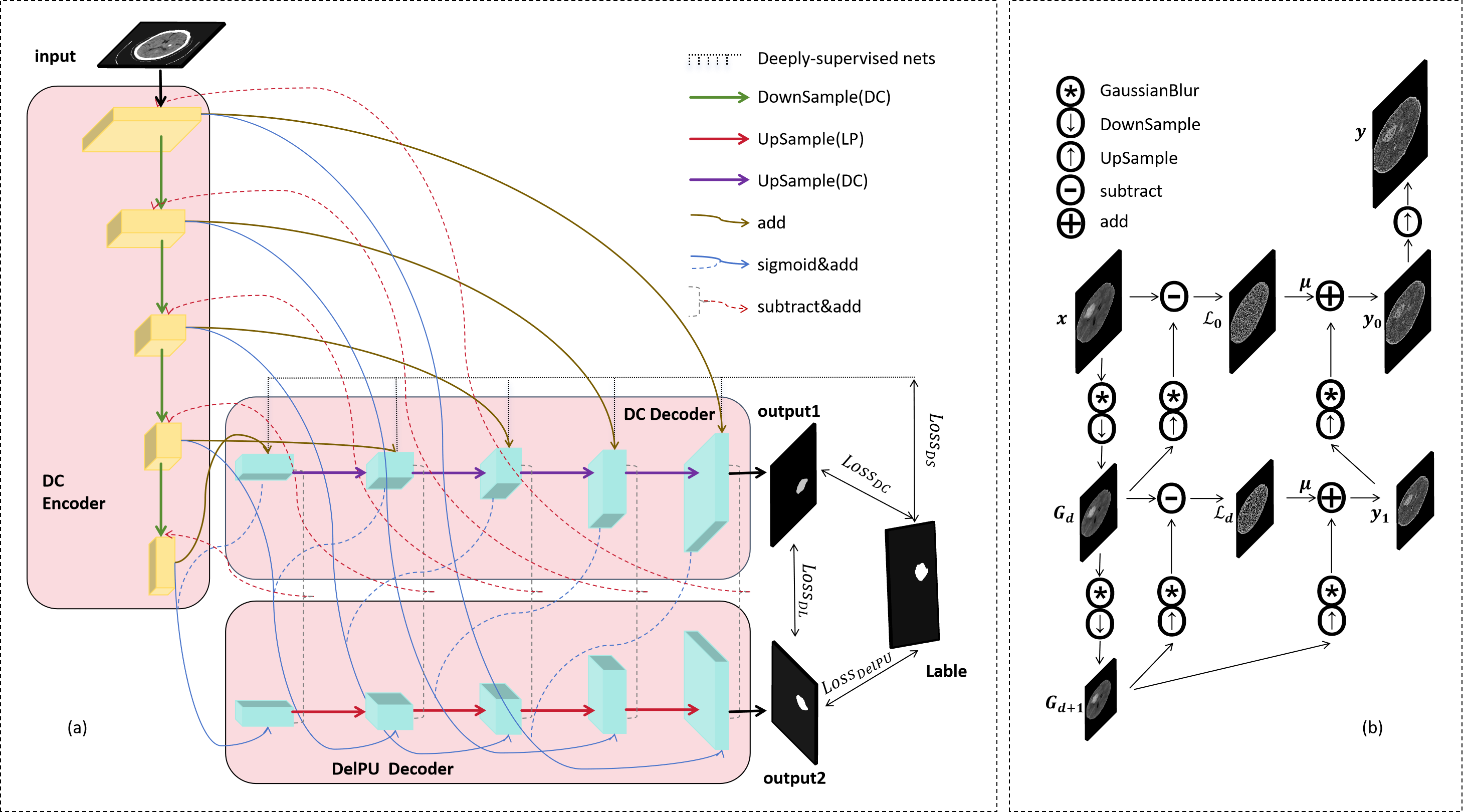} 
    \caption{(a) Network architecture of SWDL-Net, highlighting its hybrid design integrating Laplacian pyramid processing and deep convolutional pathways. (b) Schematic diagram illustrating the principle of deep laplacian pyramid upsampling, emphasizing its role in multi-resolution feature refinement.} 
    \label{fig:1} 
\end{figure*}

\normalcolor  
As illustrated in Fig.~\ref{fig:1}(a), the core principles of SWDL-Net are grounded in three crucial steps: Firstly, a DC encoder is utilized to accomplish feature extraction. Secondly, a dual-decoder architecture generates the difference between feature maps at each hierarchical stratum. Lastly, the model recursively learns the discriminative information contained within this difference, specifically focusing on the disparities between the deep Laplacian pyramid upsampling method and the deep convolutional upsampling technique.

\vspace{5mm}

\subsection{SWDL: \underline{S}tratum-\underline{W}ise \underline{D}ifference \underline{L}earning}
To harness the discriminative information embedded in the stratum-wise differences between the two decoders, we employ distinct upsampling strategies: one decoder utilizes deep transposed convolutions (DC decoder), while the other adopts deep Laplacian pyramid upsampling (DelPU decoder).

To further enrich the feature diversity, we optimize each decoder with different loss functions: the DC decoder employs cross-entropy (CE) loss for voxel-wise accuracy, while the DelPU decoder uses Dice loss to emphasize region-wise consistency. Although both decoders share the same segmentation objective on labeled data, their distinct learning dynamics foster complementary feature representations.

The hyperparameter $T \in \mathbb{Z}^+$ controls the iteration period in difference learning. For a given stratum $s$ at iteration phase $p \in \{1,...,T\}$, the feature-stratum difference $\Delta^{(s,p)}$ is computed as:
\begin{equation}
    \Delta^{(s,p)} = f_{\theta_{\mathrm{DC}}}^{s}(y_{\mathrm{DC},p}^{s+1}) - f_{\theta_{\mathrm{DelPU}}}^{s}(y_{\mathrm{DelPU},p}^{s+1})
\end{equation}
where $y_{\mathrm{DC},p}^{s+1}$ and $y_{\mathrm{DelPU},p}^{s+1}$ represent the decoder features from stratum $s+1$ produced by the DC decoder $f_{\theta_{\mathrm{DC}}}^{s+1}$ and DelPU decoder $f_{\theta_{\mathrm{DelPU}}}^{s+1}$, respectively, at the $p$-th iteration.

The inherent discrepancy between these feature representations provides valuable supplementary information for learning from unlabeled data. We propose integrating these discrepancies into the encoder at subsequent iterations through:
\begin{equation}
    y_{E,p}^{s} = 
    \begin{cases}
        f_{\theta_{E}}^{s}(x), & s = 1 \\
        f_{\theta_{E}}^{s}(y_{E,p}^{s-1}), & s > 1 \text{ and } p = 1 \\
        f_{\theta_{E}}^{s}(y_{E,p}^{s-1} + \xi \Delta^{(s-1,p-1)}), & s > 1 \text{ and } p > 1
    \end{cases}
\end{equation}
where $\xi$ is a hyperparameter controlling the discrepancy influence, and $y_{E,p}^{s-1}$ denotes the feature embedding from stratum $s-1$ generated by the encoder $f_{\theta_{E}}^{s}$ at the current $p$-th iteration.

\subsection{DelPU: \underline{De}ep \underline{L}aplacian \underline{P}yramid \underline{U}psampling}
The DelPU module is designed to preserve high-frequency details through multi-scale feature representation. As illustrated in Fig.~\ref{fig:1}(b), our framework implements a four-stage hierarchical processing pipeline:

\begin{enumerate}
    \item \textbf{Gaussian pyramid construction}: The input feature map $x$ is progressively downsampled to form Gaussian pyramid stratum:
    \begin{equation}
        G_d = \begin{cases}
            x, & d = 0 \\
            \text{DW}\left(G_{d-1} \otimes g\right), & d > 0
        \end{cases}
        \label{eq:gaussian_pyramid}
    \end{equation}
    where $\text{DW}(\cdot)$ denotes the downsampling operation implemented via trilinear interpolation, which computes weighted averages using the 8 nearest neighboring points. The Gaussian kernel is represented by $g$, and $G_d$ corresponds to the output Gaussian image at stratum depth $d$.

    \item \textbf{Laplacian pyramid generation}: High-frequency components are extracted through inter-level differential operations:
    \begin{equation}
        \mathcal{L}_d = G_d - \text{UP}\left(G_{d+1}\right) \otimes g
        \label{eq:laplacian_pyramid}
    \end{equation}
    where $\text{UP}(\cdot)$ represents the upsampling operation, also implemented using trilinear interpolation. This operation first maps each pixel of the original image to a $2\times2\times2$ region in the output image, with newly created positions filled through weighted averaging of the 8 nearest neighbors. $\mathcal{L}_d$ denotes the output Laplacian image at level $d$.

    \item \textbf{Weighted reconstruction}: Laplacian components are adaptively enhanced to emphasize salient features:
    \begin{equation}
    y_d = \begin{cases}
        \mu \mathcal{L}_d + UP\left(y_{d+1}\right) \otimes g, \quad \quad d < \mathcal{D} \\
        G_d, \quad \quad d = \mathcal{D}
    \end{cases}
    \label{eq:reconstruction}
    \end{equation}
    where $y_d$ represents the reconstructed image at stratum depth $d$, $\mu$ is a hyperparameter whose magnitude influences the strength of edge sharpening capability during the upsampling process of the Laplacian pyramid. A larger value of $\mu$ enhances the edge sharpening capability, whereas a smaller value weakens it.

    \item \textbf{Final upsampling}: The reconstructed features $y_0$ are upscaled to the target resolution, yielding the final output image $y$:
\end{enumerate}

The pyramid stratum depth $\mathcal{D}$ critically determines the upsampling performance. While deeper pyramids (with more strata) enable finer multi-scale decomposition, they introduce three fundamental trade-offs: (1) quadratic growth in computational complexity, (2) potential contamination by redundant low-frequency components, and (3) degenerate feature representations when the base resolution becomes insufficient (typically below $2\times2\times2$ voxels).

To address these limitations, we propose a stratum-adaptive algorithm that dynamically optimizes the pyramid depth according to input dimensions (see Equation~\ref{eq:depth_selection}). This mechanism achieves an optimal balance between reconstruction fidelity and computational efficiency while preventing information redundancy through dimensional analysis of feature maps at each scale:

\begin{equation}
    \mathcal{D} = \begin{cases}
        0, & \text{max\_dim}(x) \leq 8 \\
        1, & 8 < \text{max\_dim}(x) \leq 32 \\
        2, & 32 < \text{max\_dim}(x) \leq 64
    \end{cases}
    \label{eq:depth_selection}
\end{equation}
where $\text{max\_dim}(\cdot)$ returns the maximum dimension of the input.

\subsection{Training and Inference}
For labeled data, we employ a deeply supervised Dice loss for the DC decoder ($f_{\theta_{\mathrm{DC}}}$) and a cross-entropy (CE) loss for the DelPU decoder ($f_{\theta_{\mathrm{DelPU}}}$), ensuring distinct feature learning for each decoder. The supervised loss function comprises three components:

\begin{equation}
    \begin{aligned}
        \mathrm{Loss}_{\mathrm{sup}} &= \mathrm{Loss}_{\mathrm{DC}} + \mathrm{Loss}_{\mathrm{DelPU}} + \mathrm{Loss}_{\mathrm{DS}}, \\
        \mathrm{Loss}_{\mathrm{DC}} &= \frac{1}{N} \sum_{i=1}^{N} \mathrm{Dice}\left(f_{\theta_{\mathrm{DC}}}(x_{i}), y_{i}\right), \\
        \mathrm{Loss}_{\mathrm{DelPU}} &= \mathrm{CE}\left(f_{\theta_{\mathrm{D}}}(x_{i}), y_{i}\right), \\
        \mathrm{Loss}_{\mathrm{DS}} &= \frac{1}{N} \sum_{i=1}^{N} \sum_{l=1}^{L} \omega_{s} \mathrm{Dice}\left(\zeta\left(f_{\theta_{\mathrm{DC}}}(x_{i}), y_{i}\right)\right).
    \end{aligned}
\end{equation}

where $N$ denotes the number of labeled samples, $\mathcal{S}$ represents the number of hierarchical stratum, $\omega_{s}$ is the balancing weight for the $s$-th stratum loss, and $\xi(\cdot)$ denotes the combination of convolution and up-sampling operations. In our experiments, we set $\omega_{s} = \{0.8, 0.6, 0.4, 0.2, 0.1\}$ for deeply supervised losses from low-stratum to high-stratum features in the DC decoder $f_{\theta_{\mathrm{DC}}}$.

For unlabeled data, we employ a difference learining loss between the DC decoder and DelPU decoder ($f_{\theta_{\mathrm{DL}}}$), and we employ mean squared error (MSE) loss to enforce consistency between decoder outputs:

\begin{equation}
    \begin{aligned}
        \mathrm{Loss}_{\mathrm{unsup}} &= \frac{1}{M} \sum_{i=1}^{M} \mathrm{MSE}\left(f_{\theta_{\mathrm{DC}}}(x_{i}), f_{\theta_{\mathrm{DelPU}}}(x_{i})\right)
    \end{aligned}
\end{equation}

where $M$ indicates the number of unlabeled samples. The overall objective function combines both supervised and unsupervised losses:

\begin{equation}
    \mathrm{Loss} = \mathrm{Loss}_{\mathrm{sup}} + \mathrm{Loss}_{\mathrm{unsup}}
\end{equation}

During inference, the model requires only a single forward pass through the encoder and primary decoder, eliminating the need for discrepancy-based learning. This efficient design facilitates practical deployment with minimal computational overhead.
\section{Materials and Evaluation methods}

\subsection{Clinical Workflow And Patient Selection}
All studies were performed under an institutional review board approved protocol at Beijing Tiantan Hospital. From April 2020 to November 2022, we retrospectively collected CT images and clinical information of a total of 99 patients with intracranial hemorrhagic  stroke admitted to the emergency room. All patients with a total of 271 cranial imaging examinations after stroke onset were included in development and validation of the proposed SWDL-Net technique.

\subsection{Data Acquisition And Image Reconstruction}

Non-contrast computed tomography (NCCT) data were acquired using a 256-slice CT scanners (Philips, GE) with voxel sizes ranging from $0.408\times 0.408\times 5$ mm$^3$ to $1\times 1\times 5$ mm$^3$.

\subsection{Data Pre-processing}

The data preprocessing stage consists of three main steps designed to enhance the quality and consistency of the input data, thereby improving the accuracy and robustness of the segmentation model.

\subsubsection{Label Generation}
The hematoma areas in the NCCT scans were manually delineated by a senior neuroradiologist with 15 years of experience using 3D Slicer \cite{pieper20043d}, a widely used open-source software for medical image analysis. This ensures high-quality and reliable annotations for training and evaluation.

\subsubsection{Resampling for Data Consistency}
To ensure spatial consistency across all volumetric images, we resampled each 3D volume to a uniform voxel spacing of $0.5\times 0.5\times 5$ mm$^3$ using trilinear interpolation. This step is crucial for standardizing the resolution of images acquired from different scanners or protocols, which may have varying voxel dimensions. Resampling not only facilitates consistent feature extraction but also ensures compatibility with the fixed input size of the deep learning model.

\subsubsection{Skull Stripping to Reduce Interference}
The skull and other non-brain tissues can introduce unnecessary noise and interfere with the segmentation of intracranial hemorrhage (ICH) regions. To address this, we applied a skull-stripping algorithm based on a combination of thresholding and morphological operations. Specifically, we first used Otsu's thresholding(150Hu) method to separate brain tissues from the background, followed by morphological closing to fill small holes and remove isolated non-brain regions \cite{wang2024task}. This step significantly reduces extraneous information, allowing the model to focus on the relevant brain structures.

\subsubsection{Extraction of Hemorrhage ROI Using Fixed Thresholding and Morphological Operations}
To improve segmentation accuracy, we extracted the region of interest (ROI) corresponding to the hemorrhage lesions. This was achieved through a two-step process:
(i) Fixed Thresholding:
We applied a fixed intensity threshold to isolate voxels with intensities indicative of hemorrhage. The threshold value(20-40Hu)  was determined empirically based on the intensity distribution of hemorrhage regions in the training data.
(ii) Morphological Refinement:
To refine the ROI, we employed morphological operations, including dilation and erosion, to smooth the boundaries of the hemorrhage regions and remove small, spurious detections. This step ensures that the extracted ROI accurately represents the hemorrhage lesions while minimizing false positives.

\subsection{Statistical Analysis}
We employed the Wilcoxon signed-rank test \cite{wilcoxon1992individual} to compare SWDL with baseline methods. For $n$ paired samples $(X_i, Y_i)$, we compute differences $D_i = X_i - Y_i$ and rank their absolute values, excluding zero differences. When ties occur in the absolute differences, we assign average ranks and include a tie correction factor $T$ in the variance calculation:

\begin{equation}
Z = \frac{W - n(n+1)/4}{\sqrt{n(n+1)(2n+1)/24 - T}} \sim \mathcal{N}(0,1)
\end{equation}

where $T = \sum (t_k^3 - t_k)/48$ for each group of $t_k$ tied ranks \cite{conover1999practical}. This adjustment maintains the test's validity when ties are present. The two-tailed p-value is derived from the standard normal distribution, providing a robust nonparametric comparison.

\subsection{Evaluation Metrics}
The performance of each method was quantitatively assessed using six standard evaluation metrics:

\begin{itemize}
    \item \textbf{Dice Similarity Coefficient (Dice)} \cite{dice1945measures}: Measures the spatial overlap between segmentation $S$ and ground truth $G$:
    \begin{equation}
        \text{Dice} = \frac{2|S \cap G|}{|S| + |G|} \in [0,1]
    \end{equation}
    where $|\cdot|$ denotes the cardinality of the set. A Dice of 1 indicates perfect overlap. This metric is widely used in medical image segmentation.
    
    \item \textbf{95\% Hausdorff Distance (HD95)} \cite{taha2015efficient}: Measures the 95th percentile of maximum surface distances between $S$ and $G$:
    \begin{equation}
        \text{HD95} = \max\{\sup_{x \in \partial S} d(x, \partial G)_{95}, \sup_{y \in \partial G} d(y, \partial S)_{95}\}
    \end{equation}
    where $d(a,B)_{95}$ denotes the 95th percentile distance. This robust variant reduces sensitivity to outliers.
    
    \item \textbf{Average Surface Distance (ASD)} \cite{heimann2009comparison}: Computes the average distance between boundaries $\partial S$ and $\partial G$:
    \begin{equation}
        \text{ASD} = \frac{1}{|\partial S| + |\partial G|} \left(
        \sum_{x \in \partial S} d(x, \partial G) + 
        \sum_{y \in \partial G} d(y, \partial S)
        \right)
    \end{equation}
    where $d(a,B)$ denotes the minimum Euclidean distance. This metric is sensitive to boundary irregularities.
    
    \item \textbf{Accuracy (Acc)} \cite{powers2020evaluation}: Measures the proportion of correctly classified voxels:
    \begin{equation}
        \text{Acc} = \frac{\text{TP} + \text{TN}}{\text{TP} + \text{TN} + \text{FP} + \text{FN}}
    \end{equation}
    where TP, TN, FP, and FN represent true positives, true negatives, false positives, and false negatives respectively.
    
    \item \textbf{Precision (Pre)} \cite{schutze2008introduction}: Evaluates the positive predictive value:
    \begin{equation}
        \text{Pre} = \frac{\text{TP}}{\text{TP} + \text{FP}}
    \end{equation}
    Critical for assessing false positive rates in medical imaging.
    
    \item \textbf{Jaccard Index (Jac)} \cite{jaccard1912distribution}: Computes the intersection over union between $S$ and $G$:
    \begin{equation}
        \text{Jac} = \frac{|S \cap G|}{|S \cup G|} = \frac{\text{Dice}}{2 - \text{Dice}}
    \end{equation}
    This metric is particularly sensitive to segmentation edges and commonly used in medical image analysis.
\end{itemize}

\subsection{Experimental Setup}
Our experimental framework builds upon established methodologies in semi-supervised medical image segmentation \cite{yu2019uncertainty,luo2022semi,zeng2024consistency}. We adopted VNet \cite{milletari2016v}  as our baseline model, consistent with previous works in this domain. The dataset partitioning was performed using the KFold method from sklearn.model\_selection, allocating 80\% of the data for training and 20\% for testing. To ensure reproducibility, we fixed the random\_state parameter to 367, guaranteeing consistent data splits across experimental iterations.

We evaluated our proposed \textbf{SWDL-Net} against multiple semi-supervised methods under the 2\% labeled data setting. For model training, we implemented comprehensive random seed control, setting seeds for Python (1337), NumPy, and PyTorch \cite{paszke1912pytorch} (both CPU and GPU operations). This rigorous approach ensures deterministic behavior across experimental runs. The optimization was performed using SGD \cite{sutskever2013importance} with a learning rate of 0.01, weight decay of 0.0001, and momentum of 0.9. Key hyperparameters included iteration times $t$ and $\lambda$, which were carefully tuned for optimal performance.

The network processed input volumes of size $16 \times 64 \times 64$ with a batch size of 32, comprising 2 labeled and 30 unlabeled cubic patches. All experiments were conducted using PyTorch \cite{paszke1912pytorch} on an NVIDIA GeForce RTX 3090 GPU. We employed comprehensive evaluation metrics including Dice coefficient, 95\% Hausdorff Distance, Average Surface Distance, accuracy, precision, and Jaccard index. Performance metrics are reported with their respective confidence intervals to provide robust statistical analysis.

\section{Results}
\subsection{Qualitative Results}
\begin{figure*}[htbp]
    \centering
    \includegraphics[width=0.95\textwidth]{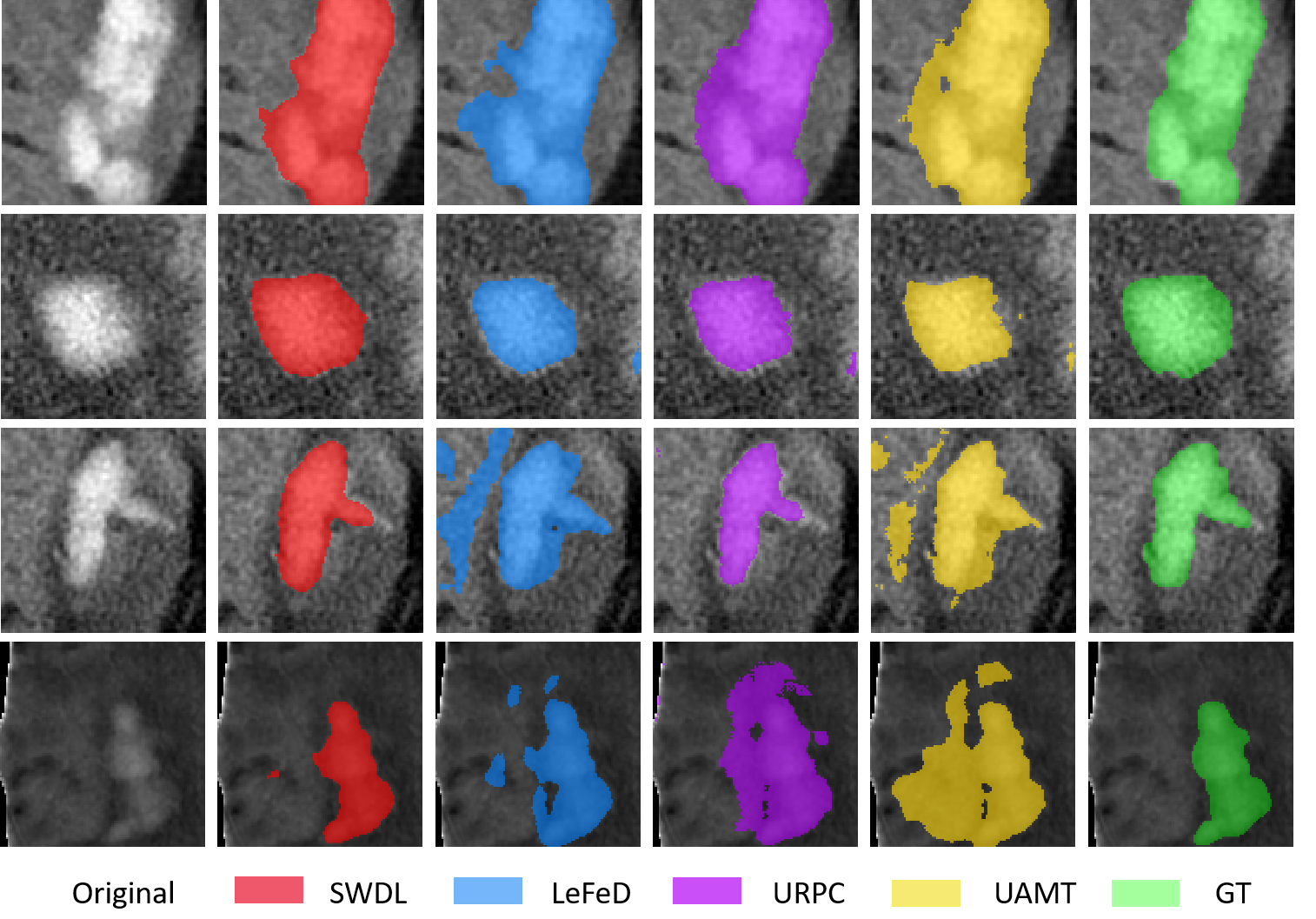}
    \caption{Comparative visualization of segmentation performance across different methods for Primary Basal Ganglia and Thalamic Hemorrhages (PBGTH). From left to right: Original CT slices, SWDL results, LeFeD results, URPC results, UAMT results and Ground truth.}
    \label{fig:2}
\end{figure*}

Fig.~\ref{fig:2} provides a visual comparison of segmentation performance across methods for challenging Primary Basal Ganglia and Thalamic Hemorrhages (PBGTH) cases. SWDL demonstrates superior anatomical fidelity and clinical relevance compared to semi-supervised baselines.

As demonstrated in the Fig.~\ref{fig:2}, our SWDL framework achieves superior segmentation performance compared to existing semi-supervised learning (SSL) approaches. The visual results clearly show that SWDL generates more anatomically plausible and clinically relevant segmentations, outperforming competing methods in three key aspects: (1) anatomical fidelity through precise hemorrhage localization, (2) boundary definition accuracy in low-contrast regions, and (3) morphological consistency with ground truth annotations. 

The visual results corroborate our quantitative findings, demonstrating SWDL's ability to produce clinically plausible segmentations that would be most useful for surgical planning and volume monitoring in clinical practice. The method particularly excels in preserving critical anatomical details that influence treatment decisions, such as core hemorrhage extension into adjacent structures and precise volume delineation.

\subsection{Quantitative Results}
As demonstrated in Table~\ref{tab:results}, our proposed SWDL framework achieves superior performance compared to existing semi-supervised methods (UAMT \cite{yu2019uncertainty}, URPC \cite{luo2022semi}, and LeFeD \cite{zeng2024consistency}) across all six evaluation metrics using only 2\% labeled data. SWDL obtains the highest Dice score (89.32\%), lowest HD95 (2.06 voxel), and best Jaccard index (80.87\%), outperforming the second-best semi-supervised method (LeFeD) by 1.79\%, 41.3\%, and 3.49\% respectively. Remarkably, SWDL achieves 96.6\% of the fully supervised VNet's \cite{milletari2016v} Dice performance while using only 2\% of the labeled data. The consistently narrow confidence intervals across all metrics (e.g., Dice CI width of 1.83 versus 2.53-5.56 for other SSL methods) demonstrate SWDL's robustness, particularly crucial for clinical applications where measurement reliability is paramount.

\begin{table*}[t]
\caption{Quantitative results on PBGTH dataset with 2\% labeled data. All values reported as mean [95\% confidence interval]. Best results highlighted in bold. Supervised VNet \cite{milletari2016v} results (100\% labels) provided as reference.}
\label{tab:results}
\centering
\resizebox{\textwidth}{!}{
\begin{tabular}{lcccccc}
\toprule
\textbf{Method} & \textbf{Dice $\uparrow$ (\%)} & \textbf{HD95 $\downarrow$ (voxel)} & \textbf{ASD $\downarrow$ (voxel)} & \textbf{Acc $\uparrow$ (\%)} & \textbf{Pre $\uparrow$ (\%)} & \textbf{Jac $\uparrow$ (\%)} \\
\midrule
UAMT \cite{yu2019uncertainty} & 84.56 [81.78,87.34] & 6.23 [4.15,8.31] & 2.05 [1.21,2.90] & 94.15 [93.33,94.97] & 82.83 [78.91,86.75] & 74.39 [71.01,77.76] \\
URPC \cite{luo2022semi} & 86.73 [85.41,88.04] & 3.61 [2.26,4.95] & 1.07 [0.73,1.41] & 94.95 [84.27,95.63] & 88.87 [87.14,90.61] & 76.88 [74.91,78.86] \\
LeFeD \cite{zeng2024consistency} & 87.53 [86.27,88.80] & 3.51 [2.47,4.54] & 1.01 [0.75,1.27] & 95.14 [94.46,95.82] & 88.24 [86.10,90.38] & 78.14 [76.21,80.06] \\
\textbf{SWDL} & \textbf{89.32 [88.40,90.23]} & \textbf{2.06 [1.58,2.54]} & \textbf{0.59 [0.45,0.73]} & \textbf{95.92 [95.38,96.46]} & \textbf{90.81 [89.51,92.12]} & \textbf{80.87 [79.39,82.35]} \\
\midrule
Supervised (VNet  \cite{milletari2016v} ) & 92.48 [91.81,93.15] & 1.15 [1.05,1.25] & 0.29 [0.25,0.32] & 97.15 [96.80,97.51] & 93.54 [92.72,94.36] & 86.11 [84.97,87.25] \\
\bottomrule
\end{tabular}
}
\end{table*}

\begin{figure*}[htbp]
\centering
\includegraphics[width=\textwidth]{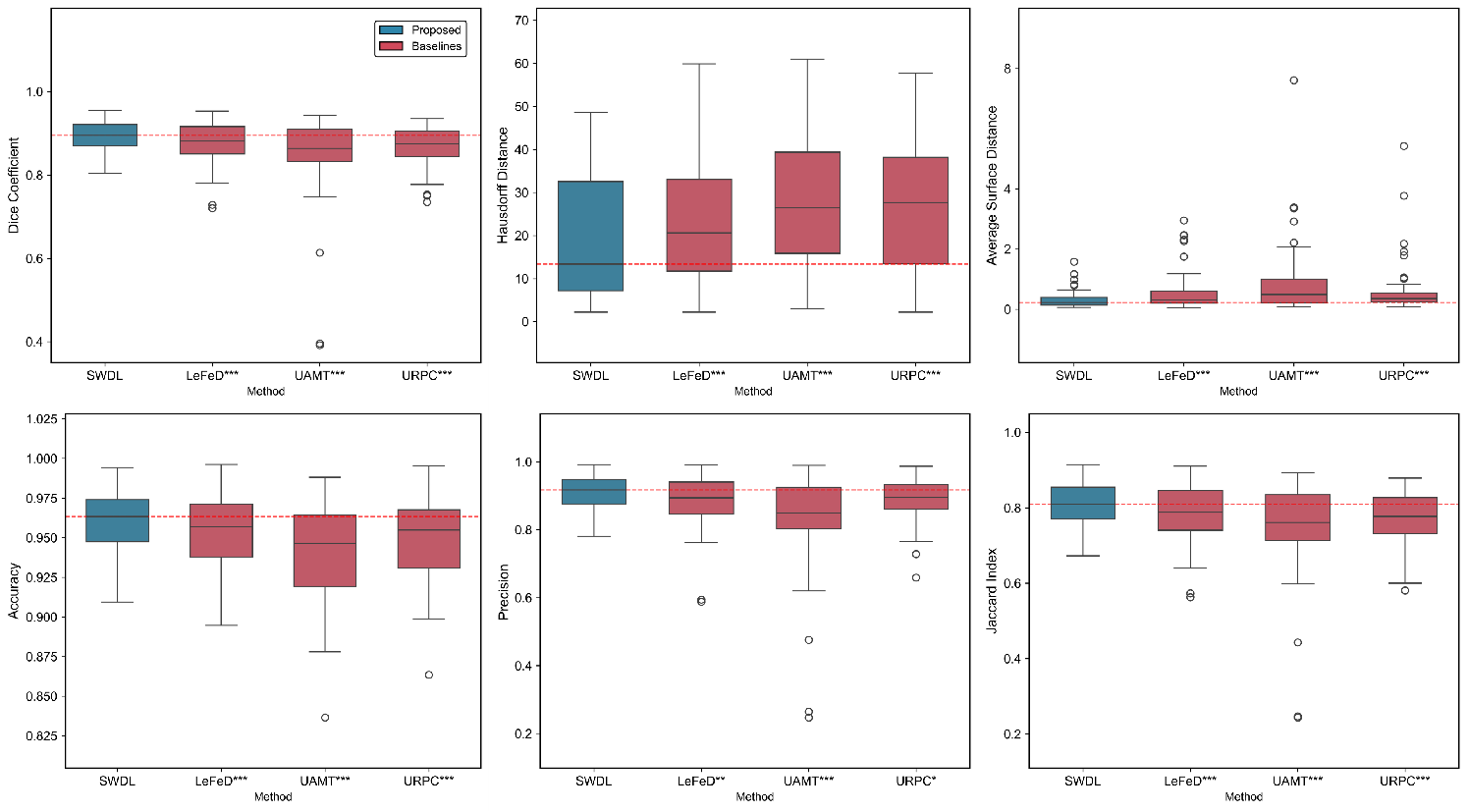}
\caption{Performance comparison of SWDL versus baseline methods across six segmentation metrics. Boxplots show distributions (median ± IQR), with dashed lines indicating SWDL's median. Significance levels: $***p<0.001$, $**p<0.01$, $*p<0.05$ (Wilcoxon test, n=55 cases).}
\label{fig:3}
\end{figure*}

To further validate the statistical significance of these improvements, Wilcoxon signed-rank tests were conducted across all evaluated metrics. Wilcoxon signed-rank tests reveal significant improvements in all metrics. Fig.~\ref{fig:3} visually confirms SWDL's statistical superiority through comprehensive boxplot analysis. The consistently higher median values (indicated by red dashed lines) and tighter interquartile ranges across all metrics demonstrate SWDL's enhanced segmentation accuracy and stability. Notably, SWDL exhibits exceptional performance in boundary-sensitive metrics, particularly in terms of Average Surface Distance and Hausdorff Distance 95\%. Specifically, SWDL achieves a significant improvement over baseline methods, with reductions in ASD ranging from 41.6\% to 71.2\% and in HD95 ranging from 41.3\% to 66.9\%. This remarkable enhancement underscores SWDL's ability to integrate the strengths of Laplacian Pyramids, which excel in detail recovery and edge sharpening, with the foundational capabilities of deep convolutional networks.

\subsection{Ablation Analysis}
\subsubsection{Ablation Study on Hyperparameter $\mu$}
$\mu$ is a hyperparameter whose magnitude influences the strength of edge sharpening capability during the upsampling process of the Laplacian pyramid. A larger value of $\mu$ enhances the edge sharpening capability, whereas a smaller value weakens it.

Our comprehensive analysis of the consistency weight hyperparameter $\mu$ demonstrates a clear performance peak at $\mu=1.5$, as evidenced by the quantitative results in Table~\ref{tab:ablation}. The $\mu=1.5$ configuration achieves statistically significant improvements across all evaluation metrics, establishing it as the optimal setting for our SWDL framework. Most notably, this setting produces the highest Dice coefficient (89.32\%, 95\% CI [88.40,90.23]), representing a 1.92\% improvement over $\mu=1.4$. 

The performance advantage becomes particularly apparent when examining the precision (90.81\%) and Jaccard index (80.87\%). This consistent superiority across diverse metrics suggests that $\mu=1.5$ achieves an ideal equilibrium between the supervised loss and the unsupervised loss. The narrow confidence intervals observed for $\mu=1.5$ (e.g., Dice CI width of 1.83) further reinforce the statistical reliability of these findings. These results collectively indicate that an excessively large value of $\mu$ ($\mu>1.5$) degrades segmentation quality, while an insufficiently small value of $\mu$ ($\mu<1.5$) fails to fully exploit the unlabeled data.

\begin{table*}[t]
\caption{Quantitative ablation study examining $\mu$ parameter variations in SWDL framework (PBGTH dataset with 2\% labeled data). All values reported as mean [95\% confidence interval]. Best results highlighted in bold.}
\label{tab:ablation}
\centering
\resizebox{\textwidth}{!}{
\begin{tabular}{lcccccc}
\toprule
\textbf{Method} & \textbf{Dice $\uparrow$ (\%)} & \textbf{HD95 $\downarrow$ (voxel)} & \textbf{ASD $\downarrow$ (voxel)} & \textbf{Acc $\uparrow$ (\%)} & \textbf{Pre $\uparrow$ (\%)} & \textbf{Jac $\uparrow$ (\%)} \\
\midrule
SWDL($\mu=1.3$) & 88.89 [87.84,89.94] & 2.22 [1.68,2.76] & 0.67 [0.52,0.83] & 95.79 [95.23,96.35] & 90.60 [89.28,91.91] & 80.22 [78.56,81.88] \\
SWDL($\mu=1.4$) & 87.64 [86.62,88.65] & 2.66 [2.09,3.23] & 0.91 [0.73,1.08] & 94.88 [94.14,95.61] & 85.94 [84.17,87.70] & 78.19 [76.61,79.77] \\
\textbf{SWDL($\mu=1.5$)} & \textbf{89.32 [88.40,90.23]} & \textbf{2.06 [1.58,2.54]} & \textbf{0.59 [0.45,0.73]} & \textbf{95.92 [95.38,96.46]} & \textbf{90.81 [89.51,92.12]} & \textbf{80.87 [79.39,82.35]} \\
SWDL($\mu=1.6$) & 88.34 [87.19,89.49] & 2.85 [1.73,3.97] & 0.93 [0.56,1.31] & 95.49 [94.89,96.10] & 88.31 [86.53,90.09] & 79.36 [77.61,81.12] \\
SWDL($\mu=1.7$) & 88.90 [87.97,89.82] & 2.28 [1.69,2.87] & 0.66 [0.51,0.81] & 95.67 [95.09,96.25] & 89.59 [88.28,90.91] & 80.18 [78.70,81.66] \\
\bottomrule
\end{tabular}
}
\end{table*}

\subsubsection{Ablation Study on Hyperparameter $T$}
The ablation study investigating the impact of hyperparameter $T$ (iteration times for discrepancy learning) reveals significant performance variations across different settings. As shown in Table~\ref{tab:T_ablation}, SWDL with $T=3$ achieves superior segmentation performance across all metrics, attaining the highest Dice score (89.32, 95\% CI [88.40,90.23]), lowest HD95 (2.06 voxel), and optimal results in ASD, Accuracy, Precision, and Jaccard index. Performance degrades when $T$ deviates from this optimal value - $T=4$ shows the most substantial performance drop, while $T=2$ and $T=5$ demonstrate intermediate results. This suggests that $T=3$ provides the ideal balance between sufficient discrepancy learning and prevention of overfitting.

\begin{table*}[t]
\centering
\caption{Quantitative ablation study examining $T$ parameter variations in SWDL framework (PBGTH dataset with 2\% labeled data). All values reported as mean [95\% confidence interval]. Best results highlighted in bold.}
\label{tab:T_ablation}
\begin{tabular}{lcccccc}
\toprule
\textbf{Method} & \textbf{Dice $\uparrow$ (\%)} & \textbf{HD95 $\downarrow$ (voxel)} & \textbf{ASD $\downarrow$ (voxel)} & \textbf{Acc $\uparrow$ (\%)} & \textbf{Pre $\uparrow$ (\%)} & \textbf{Jac $\uparrow$ (\%)} \\
\midrule
SWDL($T=2$) & 89.07 [88.15,90.00] & 2.42 [1.86,2.97] & 0.73 [0.56,0.90] & 95.78 [95.21,96.34] & 90.33 [89.02,91.65] & 80.47 [78.98,81.96] \\
\textbf{SWDL($T=3$)} & \textbf{89.32 [88.40,90.23]} & \textbf{2.06 [1.58,2.54]} & \textbf{0.59 [0.45,0.73]} & \textbf{95.92 [95.38,96.46]} & \textbf{90.81 [89.51,92.12]} & \textbf{80.87 [79.39,82.35]} \\
SWDL($T=4$) & 87.75 [86.36,89.13] & 3.51 [2.43,4.59] & 1.03 [0.74,1.31] & 95.23 [94.49,95.96] & 88.33 [86.06,90.60] & 78.52 [76.47,80.58] \\
SWDL($T=5$) & 88.56 [87.74,89.37] & 2.06 [1.75,2.38] & 0.82 [0.67,0.97] & 95.39 [94.80,95.98] & 86.92 [85.53,88.32] & 79.60 [78.28,80.91] \\
\bottomrule
\end{tabular}
\end{table*}

\subsubsection{Ablation Study on Hyperparameter $\xi$}
The hyperparameter $\xi$ serves as a critical weighting factor to balance the influence of discrepancy in our SWDL framework. Through systematic evaluation across different $\xi$ values (from $1\times10^{-2}$ to $1\times10^{-4}$), we observe significant performance variations as shown in Table~\ref{tab:zeta_ablation}. The SWDL framework achieves its optimal segmentation performance at $\xi=1\times10^{-3}$, demonstrating superior results across all metrics. Specifically, this configuration yields the highest Dice score (89.32\%), lowest HD95 (2.06 mm), and best ASD (0.59 mm), along with peak performance in accuracy (95.92\%), precision (90.81\%), and Jaccard index (80.87\%). Both larger ($\xi=1\times10^{-2}$) and smaller ($\xi=1\times10^{-4}$) values lead to notably degraded performance, confirming that $\xi=1\times10^{-3}$ represents the ideal balance for our framework.

\begin{table*}[t]
\centering
\caption{Quantitative ablation study examining $\xi$ parameter variations in SWDL framework (PBGTH dataset with 2\% labeled data). All values reported as mean [95\% confidence interval]. Best results highlighted in bold.}
\label{tab:zeta_ablation}
\resizebox{\textwidth}{!}{
\begin{tabular}{lcccccc}
\toprule
\textbf{Method} & \textbf{Dice $\uparrow$ (\%)} & \textbf{HD95 $\downarrow$ (voxel)} & \textbf{ASD $\downarrow$ (voxel)} & \textbf{Acc $\uparrow$ (\%)} & \textbf{Pre $\uparrow$ (\%)} & \textbf{Jac $\uparrow$ (\%)} \\
\midrule
SWDL ($\xi=1\times10^{-2}$) & 88.10 [87.11, 89.08] & 2.75 [2.19, 3.32] & 0.91 [0.73, 1.09] & 95.38 [94.80, 95.96] & 88.35 [86.80, 89.90] & 78.92 [77.36, 80.47] \\
\textbf{SWDL ($\xi=1\times10^{-3}$)} & \textbf{89.32 [88.40, 90.23]} & \textbf{2.06 [1.58, 2.54]} & \textbf{0.59 [0.45, 0.73]} & \textbf{95.92 [95.38, 96.46]} & \textbf{90.81 [89.51, 92.12]} & \textbf{80.87 [79.39, 82.35]} \\
SWDL ($\xi=1\times10^{-4}$) & 86.63 [85.02, 88.23] & 4.04 [2.81, 5.28] & 1.25 [0.89, 1.61] & 94.47 [93.55, 95.40] & 84.42 [81.61, 87.22] & 76.86 [74.59, 79.12] \\
\bottomrule
\end{tabular}
}
\end{table*}

\subsubsection{Component Analysis}
To comprehensively evaluate the contribution of each component in our proposed SWDL framework, we conducted systematic ablation studies by progressively removing key modules. The full SWDL architecture integrates two critical components: Deep Supervision (DS) and Difference Learning (DL). We examined three ablated variants: (1) SWDL-DS removes the Deep Supervision module which enhances feature learning at multiple scales; (2) SWDL-DL eliminates the Difference Learning component that captures subtle feature variations; and (3) SWDL-DS-DL represents the baseline model without both critical components. This systematic decomposition allows us to quantify each module's contribution to the overall segmentation performance.

\begin{table*}[t]
\centering
\caption{Quantitative Component Analysis ablation study examining the impact of configurations involving DL and DS in the SWDL framework on the PBGTH dataset with 2\% labeled data. All values reported as mean [95\% confidence interval]. Best results highlighted in bold. (DL: Difference Learning; DS: Deep Supervision)}
\label{tab:ablation2}
\resizebox{\textwidth}{!}{
\begin{tabular}{lcccccc}
\toprule
\textbf{Method} & \textbf{Dice $\uparrow$ (\%)} & \textbf{HD95 $\downarrow$ (voxel)} & \textbf{ASD $\downarrow$ (voxel)} & \textbf{Acc $\uparrow$ (\%)} & \textbf{Pre $\uparrow$ (\%)} & \textbf{Jac $\uparrow$ (\%)} \\
\midrule
SWDL & \textbf{89.32 [88.40,90.23]} & \textbf{2.06 [1.58,2.54]} & \textbf{0.59 [0.45,0.73]} & \textbf{95.92 [95.38,96.46]} & \textbf{90.81 [89.51,92.12]} & \textbf{80.87 [79.39,82.35]} \\
SWDL-DS & 88.82 [87.92,89.72] & 2.65 [1.83,3.48] & 0.87 [0.68,1.07] & 95.67 [95.12,96.22] & 89.53 [88.20,90.85] & 80.05 [78.60,81.50] \\
SWDL-DL & 88.15 [87.03,89.26] & 2.61 [1.99,3.23] & 0.79 [0.62,0.96] & 95.45 [94.86,96.03] & 88.39 [86.50,90.27] & 79.05 [77.31,80.79] \\
SWDL-DS-DL & 87.00 [85.51,88.49] & 3.58 [2.45,4.70] & 1.19 [0.93,1.45] & 95.04 [94.46,95.63] & 84.89 [82.39,87.38] & 77.38 [75.26,79.50] \\
\bottomrule
\end{tabular}
}
\end{table*}

As demonstrated in Table~\ref{tab:ablation2}, the complete SWDL framework achieves superior performance across all metrics compared to its ablated versions. The full model attains a Dice score of 89.32 [88.40,90.23], significantly outperforming SWDL-DS (88.82), SWDL-DL (88.15), and SWDL-DS-DL (87.00). Similar performance gaps are observed in other metrics, particularly in HD95 where SWDL shows a 22.6\% improvement over SWDL-DS (2.06 vs 2.65) and a 42.5\% improvement over SWDL-DS-DL (2.06 vs 3.58). The consistent performance degradation when removing either or both components validates their complementary roles in the framework. The confidence intervals, further support the statistical significance of the differences between the full model and its ablated version for key metrics.

\subsection{Validation on Public Dataset}
\begin{figure*}[htbp]
\centering
\includegraphics[width=0.92\textwidth]{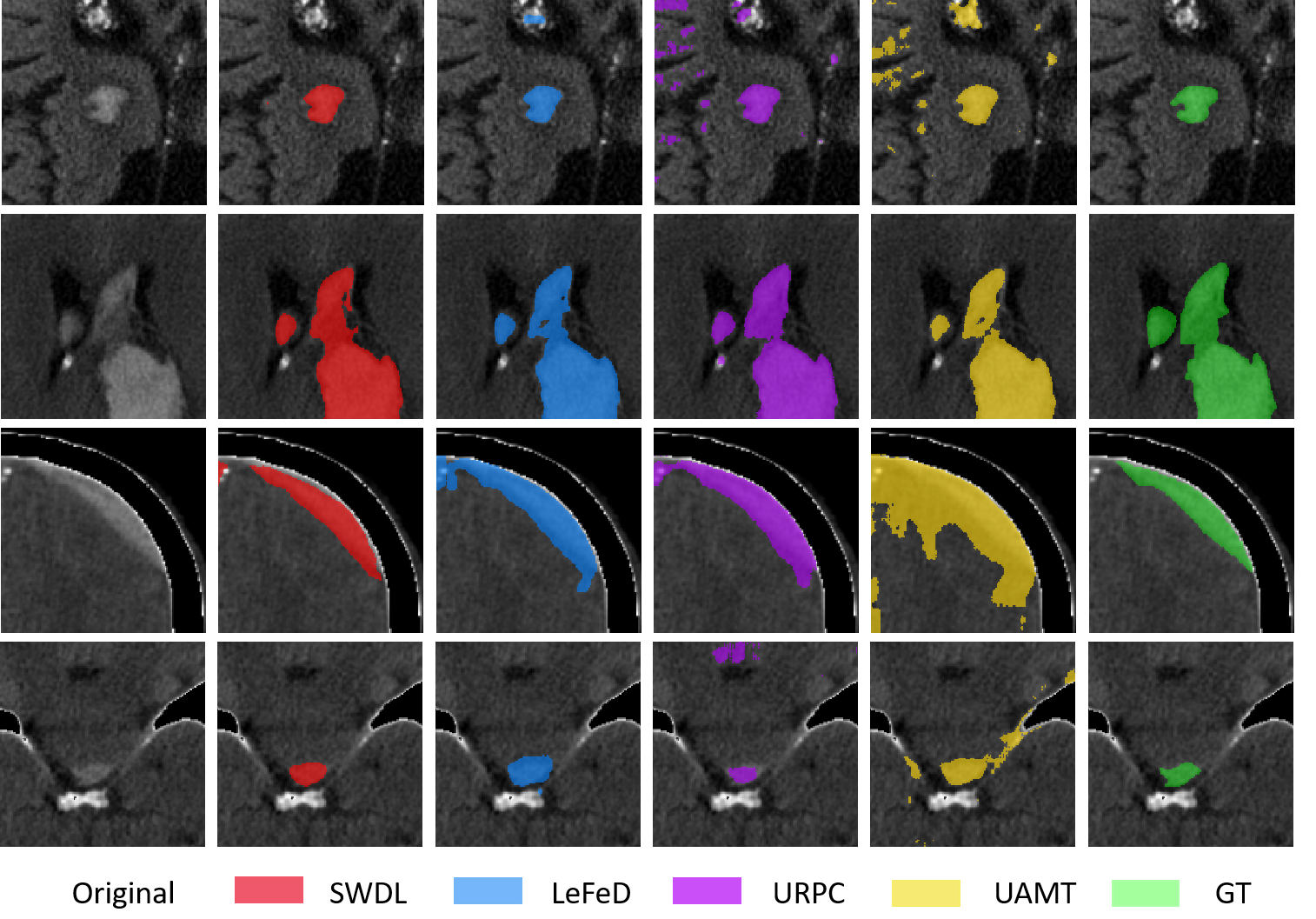}
\caption{Comparative visualization of segmentation performance across different methods for the Brain Hemorrhage Segmentation Dataset (BHSD) 2024. From left to right: Original CT slices, SWDL results, LeFeD results, URPC results, UAMT results and Ground truth.}
\label{fig:bhsd_qual}
\end{figure*}

To demonstrate generalizability, we evaluated SWDL on the Brain Hemorrhage Segmentation Dataset (BHSD) 2024 - a publicly available benchmark containing 192 non-contrast CT (NCCT) scans with expert annotations. 
We partitioned the dataset using KFold method from sklearn.model\_selection, allocating 80\% (153 cases) for training and 20\% (39 cases) for testing. Results using only 5\% labeled data (8 cases) are shown in Fig.~\ref{fig:bhsd_qual} and Table~\ref{tab:bhsd_quant}.

\begin{table*}[t]
\caption{Quantitative results on BHSD dataset with 5\% labeled data. All values reported as mean [95\% confidence interval]. Best results highlighted in bold. Supervised VNet  \cite{milletari2016v} results (100\% labels) provided as reference.}
\label{tab:bhsd_quant}
\centering
\resizebox{\textwidth}{!}{
\begin{tabular}{lcccccc}
\toprule
\textbf{Method} & \textbf{Dice $\uparrow$ (\%)} & \textbf{HD95 $\downarrow$ (voxel)} & \textbf{ASD $\downarrow$ (voxel)} & \textbf{Acc $\uparrow$ (\%)} & \textbf{Pre $\uparrow$ (\%)} & \textbf{Jac $\uparrow$ (\%)} \\
\midrule
UAMT \cite{yu2019uncertainty} & 48.70 [41.92,95.90] & 27.38 [20.47,34.29] & 10.48 [6.74,14.21] & 96.99 [96.37,97.61] & 54.00 [44.02,63.98] & 37.32 [28.78,45.85] \\
URPC \cite{luo2022semi} & 48.17 [38.73,57.60] & 27.09 [20.49,33.69] & 8.43 [5.06,11.80] & 96.50 [95.80,97.19] & 50.77 [41.23,60.30] & 36.81 [28.46,45.17] \\
LeFeD \cite{zeng2024consistency} & 50.70 [41.35,60.05] & 23.91 [17.24,30.58] & 6.85 [4.47,9.24] & 97.03 [96.39,97.66] & 59.39 [49.61,69.17] & 39.29 [30.60,47.97] \\
\textbf{SWDL} & \textbf{51.81} [42.39,61.24] & \textbf{19.78} [13.45,26.11] & \textbf{5.62} [3.30,7.95] & \textbf{97.40} [96.83,97.97] & \textbf{63.69} [54.31,73.07] & \textbf{40.44} [31.65,49.24] \\
\midrule
Supervised (VNet  \cite{milletari2016v} ) & 61.00 [52.81,69.18] & 20.37 [14.32,26.42] & 7.00 [14.32,26.42] & 97.94 [97.55,98.32] & 66.97 [58.94,75.01] & 48.59 [40.41,56.77] \\
\bottomrule
\end{tabular}
}
\end{table*}

\textbf{Qualitative Analysis:} Fig.~\ref{fig:bhsd_qual} visually demonstrates SWDL's superior segmentation performance compared to other semi-supervised methods. Our approach consistently produces results that more accurately match the ground truth in terms of hemorrhage shape, size, and location. The competing methods show varying degrees of over-segmentation, while SWDL maintains better balance between sensitivity and specificity.

\textbf{Quantitative Analysis:} As shown in Table~\ref{tab:bhsd_quant}, SWDL achieves the best performance across all metrics among semi-supervised methods. Specifically, it obtains the highest Dice score (51.81\%), Jaccard index (40.44\%), and accuracy (97.40\%), while achieving the lowest HD95 (19.78 voxels) and ASD (5.62 voxels). Notably, SWDL's precision (63.69\%) significantly outperforms other methods by 7.2-25.5\%, demonstrating its effectiveness in reducing false positives. Compared to fully supervised VNet  \cite{milletari2016v} using 100\% labeled data, SWDL achieves 85\% of its Dice performance while using only 5\% annotations, highlighting its annotation efficiency. The narrower confidence intervals across all metrics further indicate SWDL's robustness in diverse clinical scenarios.

\section{Discussion}

\subsection{Technical Advancements and Clinical Implications}
The proposed SWDL-Net represents a significant advancement in semi-supervised intracerebral hemorrhage (ICH) segmentation by addressing two critical limitations of current approaches. First, our hybrid architecture successfully reconciles the complementary strengths of Laplacian pyramid processing (for excellent edge recovery) and deep convolutional upsampling (for learnable feature enhancement), thereby overcoming the anisotropic resolution challenges inherent in non-contrast computed tomography (NCCT) data. Second, the difference learning mechanism offers a novel solution to the feature redundancy problem in multi-path architectures, enabling more efficient utilization of unlabeled data.

Clinically, SWDL's performance holds important implications for emergency stroke management. The method's robustness in low-label regimes (with 2-5\% annotations) suggests potential applicability in resource-limited settings where expert annotations are scarce. Particularly noteworthy is SWDL's superior performance in boundary-sensitive metrics (with 41.3-71.2\% improvement in Hausdorff Distance 95\% (HD95) and Average Surface Distance (ASD)), as hemorrhage volume and shape accuracy directly influence surgical planning decisions.

\subsection{Comparative Analysis with State-of-the-Art}

Our results demonstrate consistent advantages over existing semi-supervised learning (SSL) approaches across multiple dimensions. Compared to consistency-based methods like UA-MT \cite{yao2019unsupervised}, SWDL reduces boundary errors by 66.9\% (HD95: 2.06 vs. 6.23) through its Laplacian pyramid pathway. Notably, SWDL achieves 96.6\% of fully supervised performance with only 2\% labels, significantly narrowing the gap between semi- and fully-supervised paradigms \cite{chen2021semi}.

The ablation studies reveal several key insights: (1) The optimal $\mu=1.5$ balances edge enhancement and noise suppression in Laplacian processing; (2) $T=3$ iterations provide sufficient discrepancy learning without overfitting; (3) The optimal value of $\xi$ is $1\times10^{-3}$, chosen to optimally balance the influence of the discrepancy; (4) Both deep supervision and difference learning contribute substantially to performance (with a combined 2.32\% Dice improvement). 

\subsection{Limitations and Future Directions}
Several limitations warrant discussion. First, although the proposed framework leverages 3D depthwise convolution and Laplacian pyramid collaboration for 3D ICH segmentation, its ability to capture inter-slice contextual information may remain limited.  Future work should explore memory-efficient transformer architectures to capture this information. Second, while SWDL demonstrates strong performance on the PBGTH and BHSD datasets, further validation across diverse hemorrhage subtypes (e.g., subarachnoid, intraventricular) is needed to ensure its generalizability. Third, all methods performed better on the PBGTH dataset than on the BHSD dataset, likely because PBGTH focuses on Primary Basal Ganglia and Thalamic Hemorrhages, where lesion regions are relatively consistent across cases. In future work, we plan to establish classification models for different ischemic stroke subtypes and train separate segmentation models for each subtype to improve segmentation quality.

\section{Conclusion}
This paper presents SWDL-Net, a novel semi-supervised learning framework for intracranial hemorrhage segmentation that innovatively combines Laplacian pyramid processing with deep convolutional upsampling through difference learning. Our comprehensive experiments demonstrate state-of-the-art performance across multiple benchmarks, with particular strengths in boundary accuracy and low-label regimes. Key innovations include:

\begin{itemize}
    \item A dual-decoder architecture that synergizes Laplacian pyramid's edge recovery with deep convolution's feature learning capabilities
    \item An adaptive difference learning mechanism that effectively harnesses inter-pathway discrepancies
    \item Clinically validated performance achieving 96.6\% of fully supervised accuracy with only 2\% labeled data
\end{itemize}

The framework's robustness has been rigorously validated through extensive experiments (271 clinical cases and public BHSD benchmark), statistical analysis, and ablation studies. SWDL-Net not only advances the technical state-of-the-art in medical image segmentation but also offers practical solutions to the critical challenge of annotation scarcity in emergency neuroimaging. The released codebase and demonstrated performance suggest strong potential for clinical translation, particularly in time-sensitive stroke diagnosis and treatment planning scenarios.


\IEEEtriggeratref{70}

\bibliographystyle{IEEEtran}

\bibliography{ich}

\end{document}